\documentstyle[preprint,aps,epsfig]{revtex}
\begin{document}
\draft
\title{Relaxation of surface charge on rotating dielectric spheres: \\
       Implications on dynamic electrorheological effects}
\author{Jones T. K. Wan$^{1}$, K. W. Yu$^{1}$ and G. Q. Gu$^{1,2}$}
\address{$^{1}$Department of Physics, The Chinese University of Hong 
Kong, 
       \\ Shatin, New Territories, Hong Kong, China \\
 $^{2}$College of Computer Engineering, University of Shanghai for 
Science 
 and Technology, \\ Shanghai 200 093, China}
\maketitle

\begin{abstract}
We have examined the effect of an oscillatory rotation of a polarized 
dielectric particle.
The rotational motion leads to a re-distribution of the polarization 
charge on the surface of the particle. We show that 
the time averaged steady-state dipole moment is along the field direction,
but its magnitude is reduced by a factor which depends on the angular
velocity of rotation. 
As a result, the rotational motion of the particle reduces the 
electrorheological effect.
We further assume that the relaxation of polarized charge is arised 
from a finite conductivity of the particle or host medium. We calculate 
the relaxation time based on the Maxwell-Wagner theory, 
suitably generalized to include the rotational motion.
Analytic expressions for the reduction factor and the relaxation time are
given and their dependence on the angular velocity of rotation will 
be discussed.
\end{abstract}
\vskip 5mm
\pacs{PACS Numbers: 83.80.Gv, 77.84.Lf, 77.22.Gm, 41.20.Cv}

\section{Introduction}
The prediction of the strength of the electrorheological (ER) effect 
is still a main concern in theoretical investigation of ER fluids 
[1--5]. An ER fluid is a suspension of highly polarized particles in an 
insulating host. The ER effect originates from the induced interaction 
between the polarized particles in an ER fluid. 
Upon the application of an intense electric field, the particles in 
ER fluid aggregate into chains and then aggregate into columns in a 
short response time [3,4]. 
The rapid field-induced transition between the fluid and solid phase 
makes this material important both for wide industrial applications 
and for experimental and theoretical investigation.

In deriving the induced interactions between particles, existing theories 
assume that the particles are at rest [6--10]. In a realistic situation, 
the fluid flow exerts force and torque on the particles, setting the 
particles in both translational and rotational motions.
For instance, the shear flow in an ER suspension exerts a toque on the 
particles, which leads to a rotational motion of the particles 
about their centers [11].
Recent experiments gave evidences that the induced forces between 
the rotating particles can be different from the values predicted 
by existing theories [12].

To gain some insight into the phenomenon, 
we have recently formulated a theoretical model, which describes the 
relaxation of polarized charge on the surface of a uniformly rotating 
particle \cite{PRE}.
We showed that the rotational motion of the particles reduces the induced
forces between the particles. We called the reduction of interparticle
forces due to the rotational motion of the particles the dynamic ER effects
\cite{PRE}.
In this work, we extend the consideration to an arbitrary rotational motion.
In particular, we will obtain the steady-state dipole moment of a rotating
sphere under a sinusoidal oscillatory shear motion.
We further assume that the relaxation of polarized charge is due to
a finite conductivity of the particle or host medium.
We will derive an analytic expression for the relaxation time.
The dependence of the reduction factor and the relaxation time on the 
angular velocity of rotation will also be calculated.

\section{Steady-state dipole moment}
Consider a dielectric sphere under the influence of an electric field 
$\vec{E}_0=E_0\hat{z}$;
its induced dipole moment is given by: $\vec{p}_0 = p_0\hat{z}$. 
Assume that it is under a rotational motion of angular velocity 
$\vec{\omega}= -\omega\hat{y}$. 
For a rotating dielectric sphere in an electric field, the rotational 
motion leads to a displacement of its polarized charges on the surface of 
sphere. As a result, there is a change of the dipole moment, described by 
$\vec{\omega} \times \vec{p}$. The surface charges also suffer from 
relaxation of various kinds, and the rate of change of the dipole moment 
is described by $-(\vec{p}-\vec{p}_0)/\tau$, where $\tau$ is a relaxation 
time. The two effects have to be balanced against each other, resulting 
in a steady state dipole moment $\vec{p}$, which deviates from the 
equilibrium dipole moment $\vec{p}_0$. Let the resultant dipole moment be 
$\vec{p} = p_x\hat{x}+p_y\hat{y}+p_z\hat{z}$.
The rate of change of the dipole moment is given by:
\begin{eqnarray}
{d \vec{p}\over dt} = 
\vec{\omega}\times\vec{p}-{1\over\tau}(\vec{p}-\vec{p}_0),
\end{eqnarray}
where the first term on the right hand side is due to the rotational 
motion and the second term is due to a relaxation process, 
in which the relaxation time $\tau$ is 
determined by the details of the relaxation process.
In component form, the  differential equation reads:
\begin{eqnarray*}
\dot{p}_x = -{p_x\over\tau}-\omega p_z, \ \ \
\dot{p}_y = -{p_y\over\tau}, \ \ \
\dot{p}_z = \omega p_x-{(p_z-p_0)\over\tau}.
\end{eqnarray*}
The equation for $p_y$ can be readily integrated to yield 
$p_y = p_{y0} e^{-J(t)}$, where $J(t) = \int_0^t dt/\tau$.
Since $\tau$ (can be time-dependent) is real and positive, 
$p_y$ vanishes as $t$ goes to infinity.
To solve the equations for $p_x$ and $p_z$, we use the complex notation: 
let $\tilde{p} = p_x+i p_z$ and $\tilde{p}_0 = i p_0$, $\tilde{p}$ can be 
found by solving the following differential equation:
\begin{equation}
{d\tilde{p}\over dt}=\left( i\omega-{1\over\tau}\right) \tilde{p}
  +{\tilde{p}_0\over\tau}.
\label{de-complex-p}
\end{equation}
With the initial condition $\tilde{p}=\tilde{p}_0$ when $t=0$,
Eq.(\ref{de-complex-p}) admits a standard solution:
\begin{equation}
\tilde{p} e^I - \tilde{p}_0 = \tilde{p}_0 \int_0^t {e^I\over\tau} dt, \ \ \
I = \int_0^t \left( {1\over\tau}-i\omega \right) dt,
\label{standard}
\end{equation}
where $I$ is the integration factor.
For a uniform rotational motion, $\omega=\omega_0$ is a constant, 
$I = {t\over\tau}-i\omega_0 t$,
Eq.(\ref{standard}) can be solved:
$$
\tilde{p}={\tilde{p}_0\over 1-i\omega_0\tau}
  \left( 1 - i\omega_0\tau e^{-t(1-i\omega_0\tau)/\tau} \right).
$$
As $t$ goes to infinity, we obtain the steady-state solution for a uniform
rotation:
\begin{equation}
\tilde{p}={\tilde{p}_0\over 1-i\omega_0\tau}.
\label{uniform}
\end{equation}
In general, analytic solution of Eq.(\ref{standard}) can be found only 
for a few simple cases and the integral must be evaluated numerically.
We concentrate on the steady state solution at a sufficiently long time 
and Eq.(\ref{standard}) can indeed be solved exactly.
By using the L'H$\hat{\rm{o}}$pital's rule, we find:
\begin{equation}
\tilde{p} = \tilde{p}_0 \lim_{t\to\infty}{e^I\over \tau e^I \dot{I}}
 = {\tilde{p}_0\over1-i\omega\tau},
\label{limit}
\end{equation}
where $\dot{I}$ denotes the time derivative of $I$. We have assumed that 
$\tau$ is real and positive but $\omega$ can be an arbitrary function of 
time. Eq.(\ref{limit}) is the general result for arbitrary rotational
motion, being of the same form as Eq.(\ref{uniform}).
However, the transient solution has to be calculated numerically.

For a dielectric sphere undergoing a simple harmonic oscillation, 
$\theta(t)=\theta_0 \sin{k t}$, the angular velocity is given by 
$\omega(t)=\dot{\theta}=\theta_0 k \cos{k t}$.
From Eq.(\ref{limit}), the steady state dipole moment is:
\begin{equation}
\tilde{p} = {\tilde{p}_0\over1-i\theta_0 k \tau \cos{k t}}
 =p_0 {i-\theta_0 k\tau \cos{k t}\over 1+\theta_0^2 k^2\tau^2 \cos^2{k t}}.
\end{equation}
The time dependence of the dipole moment is still periodic. 
Note that although the sphere is undergoing a simple harmonic 
oscillation, 
the dipole moment does not exhibit a simple harmonic motion.
If $\tau$ is independent of time,
we can calculate the time average of the dipole moment:
\begin{equation}
{\langle p_x \rangle\over p_0} = 0 \quad{\rm and} \quad
 {\langle p_z \rangle\over p_0} = {1\over\sqrt{1+\theta_0^2 k^2 \tau^2}}.
\label{av}
\end{equation}
As a result, the motion of particles reduces the ER effect.
We define the reduction factor $R$ as:
\begin{equation}
R={\langle p_z \rangle\over p_0}={1\over\sqrt{1+\theta_0^2 k^2 \tau^2}}.
\end{equation}
The reduction is even more significant at high frequencies, 
$R \approx 1/\theta_0 k \tau$.

\section{Calculation of relaxation time}
So far, our proposed relaxation time has no explicit expression.
If the relaxation process is originated from a finite conductivity of 
the particle or host medium, then we can calculate the relaxation time 
based on the Maxwell-Wagner theory of leaky dielectrics \cite{MW}.
For a (nonrotating) spherical inclusion embedded in a host medium, 
the expression is 
\begin{equation}
\tau=\epsilon_0 \left ( {\epsilon_1 + 2 \epsilon_m\over 
 \sigma_1 + 2 \sigma_m}\right ),
\label{MW-non-rotate}
\end{equation}
where $\epsilon_1, \epsilon_m$ ($\sigma_1, \sigma_m$) denote the 
dielectric constant (conductivity) of the sphere and 
host medium respectively, $\epsilon_0$ is the permittivity of free space.
For typical values of the permittivities and conductivities of common ER 
fluids, the relaxation time ranges from microseconds to milliseconds, 
and the dynamic ER effect can be observed in experiments. 

In order to account for the impact of a rotational motion on the 
relaxation time, we first replace 
$\epsilon_1$ in Eq.(\ref{MW-non-rotate}) by $\epsilon_1=1+\chi_1$,
where $\chi_1$ is the susceptibility of the sphere.
We already showed that the dipole moment is reduced by a factor $R$.
If we assume that the polarization is uniform throughout the sphere, 
which can be achieved when the oscillating frequency is high, 
we may write $\epsilon_1=1+R\chi_1$ in Eq.(\ref{MW-non-rotate}). 
Physically it means that the effective polarization of the sphere is 
reduced as a result of the rotational motion, leading to a reduction of 
the effective dielectric constant of the sphere. After some 
simplifications, we obtain 
\begin{equation} \tau = \tau_\infty + {\tau_0-\tau_\infty \over 
  \sqrt{1+\theta_0^2 k^2 \tau^2}},
\label{micro-relaxation-time}
\end{equation}
where
$$
\tau_\infty = \epsilon_0 \left ( {1 + 2 \epsilon_m\over 
  \sigma_1 + 2 \sigma_m}\right )
\quad{\rm and}\quad
\tau_0 = \epsilon_0 \left ( {\epsilon_1 + 2 \epsilon_m\over 
 \sigma_1 + 2 \sigma_m}\right ).
$$
It can be shown that
$\tau = \tau_0$ for $k\theta_0=0$ and $\tau \to \tau_\infty$ for 
$k\theta_0\to \infty$. 
Eq.(\ref{micro-relaxation-time}) is a self-consistent equation for $\tau$
and we can calculate the relaxation time self-consistently.

\section{Numerical results}
To examine the dependence of the reduction factor $R$ on the angular
velocity $k\theta_0$, we plot $R$ vs. $k\theta_0$ in Fig.\ref{reduction} 
for several different values of $\tau_0$. 
Without loss of generality (i.e., in terms of some unit relaxation time), 
we choose $\tau_\infty=1$ and $\tau_0 = 2, 4$ and 8 respectively.
The reduction factor decreases rapidly with the increase of $k\theta_0$, 
which means that the dipole moment is greatly reduced when both the 
oscillation frequency $k$ and the oscillation amplitude $\theta_0$ 
become large.

Next we see how the relaxation time depends on $k\theta_0$.
From Eq.(\ref{micro-relaxation-time}), $\tau$ is bounded between $\tau_0$
and $\tau_\infty$.
The lower bound $\tau_\infty$ is reached when $k\theta_0$ tends to infinity,
which is achieved at large frequency.
For a larger value of $\tau_0$, the relaxation time decreases more 
rapidly 
with $k\theta_0$.
The condition of high oscillation frequency reads:
\begin{equation}
k\gg {1\over \tau_\infty}.
\label{high-freq-cond}
\end{equation}

The time evolution of the dipole moment is worth studying. 
In Fig.\ref{px-pz-k}, we plot the steady-state solution of the 
perpendicular component $p_x/p_0$ and
parallel component $p_z/p_0$ against time.
We set $\tau_0=2$ and $\tau_\infty=1$,
$\tau$ is then calculated from Eq.(\ref{micro-relaxation-time}).
We first set $k=1$ and vary the oscillation amplitude $\theta_0$.
In the left panel of Fig.\ref{px-pz-k}, we plot $p_x/p_0$ and $p_z/p_0$
against time for $\theta_0=\pi/4$, $\theta_0=\pi/2$ and $\theta_0=\pi$
respectively. For each value of $\theta_0$, the magnitude of the 
perpendicular component $(p_x/p_0)$ has a maximum value of 0.5 
and it has a local minimum at $t=\pi/k$.
When the oscillation amplitude increases, the local minimum value decreases,
showing a large variation of $p_x/p_0$.
Next we concentrate on the parallel component $(p_z/p_0)$.
When the oscillation amplitude increases,
$p_z/p_0$ reduces in general, although the maximum value is always equal 
to unity.
This results are expected from Eq.(\ref{av}):
$$
{\langle p_x \rangle\over p_0} = 0 \quad{\rm and}
\quad{\langle p_z \rangle\over p_0} = {1\over\sqrt{1+\theta_0^2 k^2 \tau^2}}.
$$
Hence, on the average, $\langle p_z \rangle/p_0$ must decrease when we 
increase the oscillation amplitude $\theta_0$.

Now we examine the case for a constant amplitude $\theta_0$ but varying 
frequency $k$.
This is a realistic case since in experiment we can hardly increase 
the amplitude but we can easily change the frequency.
In the right panel of Fig.\ref{reduction}, we choose $\theta_0 = \pi/4$ 
and $k=1$ and 3.
The results for constant $k$ and constant $\theta_0$ show similar 
time dependence.
However, $p_x$ and $p_z$ show a larger variation in their magnitudes, 
if we increases the value of $k$.
It should be remarked that we have assumed that the oscillation frequency 
is high so that the relaxation time is nearly constant during the motion.

\section{Discussion and conclusion}
Here a few comments on our results are in order.
As our steady-state solution is general, one can extend the calculations 
to an arbitrary rotational motion.
We have shown that the motion of particles reduces the strength of the 
dipole 
moment. It is natural to further calculate the interparticle force 
between 
two rotating spheres. We expect that the interparticle force will be 
reduced substantially because the force between parallel dipoles changes 
from attractive to repulsive when their orientation varies from the 
transverse to longitudinal field case.

So far, our derivation of relaxation time is based on the mean field theory.
We may extend the Maxwell-Wagner theory to the polarization relaxation 
of oscillating particles. In this case, we should add a term 
$\rho_P \vec{v}$ to the polarization current density, where $\rho_P$ is 
the polarized charge density and $\vec{v}=\vec{\omega} \times \vec{r}$ 
is the rotating velocity. 
However, it is not possible to convert the extra term into a dielectric 
constant and the generalization becomes more complicated due to 
the nonuniform polarized charge density inside the rotating 
spherical inclusions. We are currently examining the solution of the 
more complicated boundary-value problem.

The flow in ER fluid may be nonsteady in usual operation situations. 
But the prevailing situation in theory and simulation of ER 
fluids is to use formulas derived with respect to a steady flow. 
To remedy this drawback, we will endeavor to develop calculation method 
for suspension hydrodynamics, 
and use it to study the interaction between particles and the oscillating 
fluid, and derive formulas of the force and torque exerted on particles 
for a suspension \cite{flow}.

In this work, hydrodynamic (HD) interaction effects have not been 
considered. However, electrorheological fluids are locally very 
concentrated suspensions and in considering dynamic effects, 
it seems that HD effects can be strong. And this is a future problem.

In conclusion, we have investigated the problem of how the dipole moment
of a dielectric sphere varies with time for an arbitrary rotational motion.
We have developed a formalism for the rotational motion of the sphere and
derived the relaxation time by using the mean field theory.
We have shown that the time averaged steady-state dipole moment is 
along the field direction,
but its magnitude is reduced by a factor which depends on the frequency 
of oscillation. As a result, the motion of particles reduces the ER effect.
We further calculate the relaxation time based on the Maxwell-Wagner theory.
The dependence of the reduction factor and the relaxation time on the 
angular velocity of rotation has also been discussed.

\section*{acknowledgement}
This work was supported by the Research Grants Council of the Hong Kong 
SAR Government under grant CUHK4284/00P. G. Q. G. acknowledges 
financial support from the Key Project of the National Natural 
Science Foundation of China under grant 19834070.
K. W. Y. thanks Professor Hong Sun for suggestion of the integration 
factor approach and for many fruitful discussions.

\begin{figure}[h]
\caption{The reduction factor $R$ (left panel) and the relaxation time 
(right panel) $\tau$ for $\tau_\infty = 1$.
The reduction factor decreases drastically for increasing $k\theta_0$.
The relaxation time reaches its minimum value at large $k\theta_0$.}
\label{reduction}
\end{figure}

\begin{figure}[h]
\caption{The reduced dipole moment $p_x/p_0$ and $p_z/p_0$ plotted as a
function of time for various frequency dependent $\tau$
($\tau_0 = 2$, $\tau_\infty = 1$).}
\label{px-pz-k}
\end{figure}

\newpage
\centerline{\epsfig{file=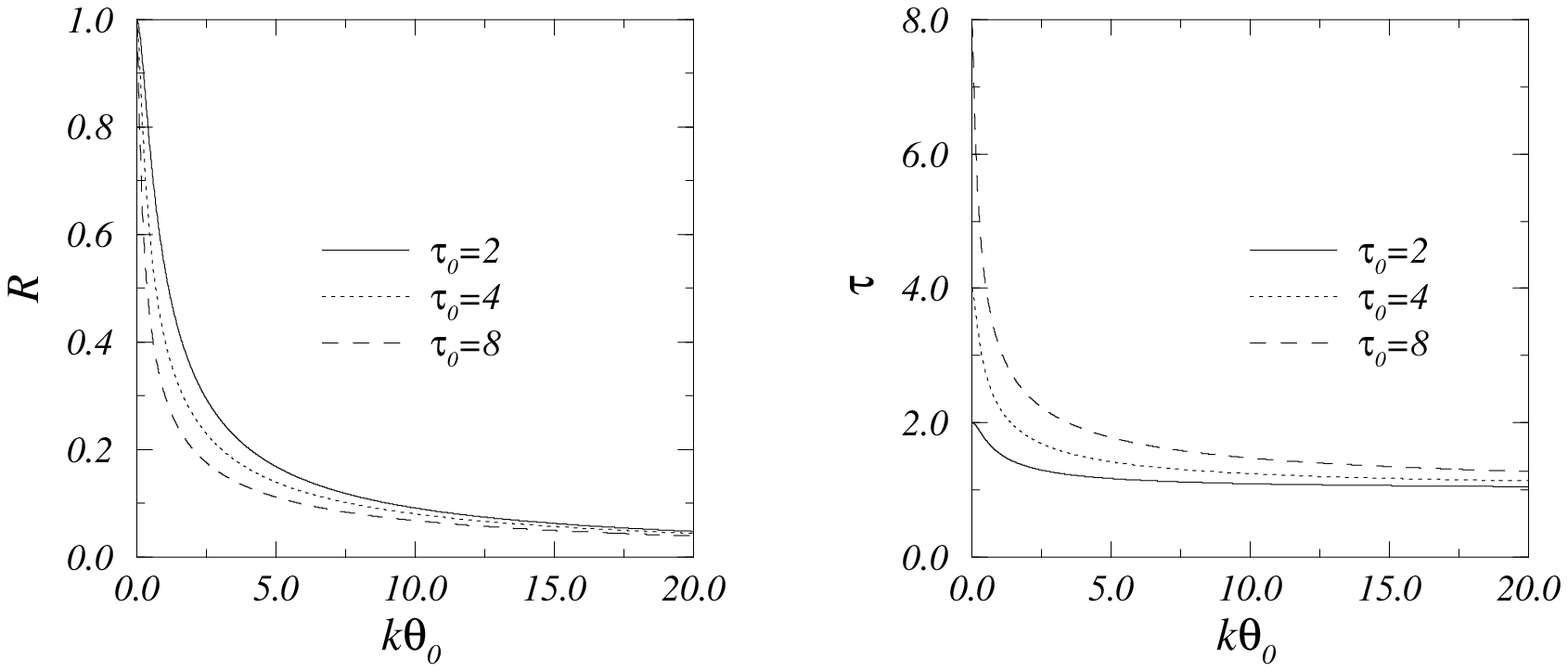,width=\linewidth}}
\centerline{Fig.1/Wan, Yu and Gu}

\newpage
\centerline{\epsfig{file=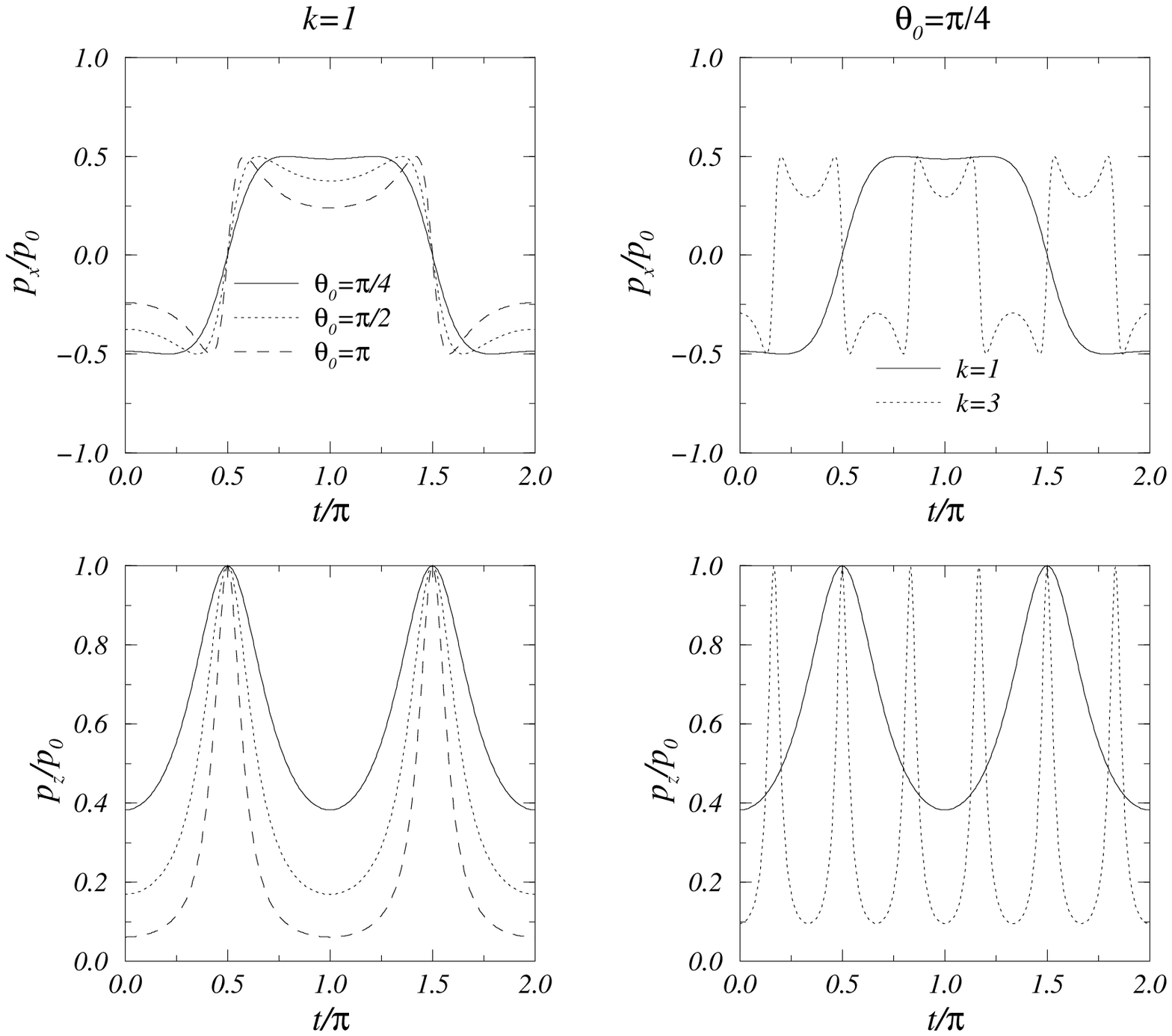,width=\linewidth}}
\centerline{Fig.2/Wan, Yu and Gu}


\begin{references}
\bibitem{1} P. P. Phul\'{e} and J. M. Ginder, MRS Bulletin {\bf 23}, 19 
 (1998).

\bibitem{2} D. J. Klingenberg, MRS Bulletin, {\bf 23}, 30 (1998).

\bibitem{3} T. C. Halsey and  W. Toor, J. Stat. Phys. {\bf 61}, 1257 (1990).

\bibitem{4} R. Tao and J. M. Sun, Phys. Rev. Lett. {\bf 67}, 398 (1991).

\bibitem{5} T. C. Halsey, Science {\bf 258}, 761 (1992).

\bibitem{6} D. J. Klingenberg, F. van Swol, and C. F. Zukoski, 
 J. Chem. Phys. {\bf 94}, 6160 (1991).

\bibitem{7} D. J. Klingenberg, F. van Swol, and C. F. Zukoski, 
 J. Chem. Phys. {\bf 91}, 7888 (1989).

\bibitem{8} D. J. Klingenberg and C. F. Zukoski, Langmuir {\bf 6}, 15 (1990).

\bibitem{9} Z. W. Wang, Z. F. Lin, and R. B. Tao, 
 Int. J. Mod. Phys. B {\bf 10}, 1153 (1996).

\bibitem{10} Z. W. Wang, Z. F. Lin, and R. B. Tao, J. Phys. D 
 {\bf 30}, 1265 (1997).

\bibitem{11} A. J. C. Ladd, J. Chem. Phys. {\bf 88}, 5051 (1988).

\bibitem{12} L. Lobry and E. Lemaire, J. Electrostat. {\bf 47}, 61 (1999).

\bibitem{PRE} Jones T. K. Wan, K. W. Yu and G. Q. Gu, Phys. Rev. E {\bf 62},
 in press (November 2000).
 
\bibitem{MW} W. B. Russel, D. A. Saville, and W. R. Schowalter,
 {\em Colloidal Dispersions} (Cambridge University Press, Cambridge,
 England, 1989).

\bibitem{flow} P. Mazur and D. Bedeaux, Physica {\bf 76}, 235 (1974).

\end{references}
\end{document}